\documentclass[3p,times,twocolumn]{elsarticle}

\usepackage{ecrc}
\usepackage{amsmath}
\usepackage{cancel}

\volume{00}

\firstpage{1}

\journalname{Nuclear Physics B Proceedings Supplement}

\runauth{}

\jid{nuphbp}

\jnltitlelogo{Nuclear Physics B Proceedings Supplement}

\usepackage{amssymb}

\usepackage[figuresright]{rotating}

\begin{document}

\begin{frontmatter}

\dochead{}

\title{An exploration of $B_s\to J/\psi\,s\bar{s}$}

\author{Robert Knegjens}
\ead{robk@nikhef.nl}

\address{Nikhef, Science Park 105, 1098 XG, Amsterdam, The Netherlands}

\begin{abstract}
Measurements of the $B_s^0$--$\bar B_s^0$ mixing phase $\phi_s$ appear to be  converging towards the SM prediction, suggesting that contributions from New Physics (NP), if present, are small.
This poses the question of whether smallish NP in $\phi_s$ can be distinguished from the hadronic uncertainties present in the interfering $B_s$ decay mode.
In this paper we discuss the potential of extracting $\phi_s$ from the decay modes $B_s\to J/\psi \, \eta^{(\prime)}$ and $B_s\to J/\psi \,f_0(980)$ and how their respective hadronic uncertainties can be controlled.
In addition, we demonstrate how the branching ratios of the former decays can be used to estimate the $\eta$--$\eta'$ mixing angles.
Finally, we point out that effective lifetime measurements of decay modes such as the $B_s\to J/\psi \,f_0(980)$ can be used to constrain both $\phi_s$ and the $B_s$ decay width difference $\Delta\Gamma_s$, complementary to the usual time-dependent tagged analysis.
\end{abstract}

\begin{keyword}
	$B_s$ decays \sep effective lifetimes \sep New Physics \sep flavour symmetry

\end{keyword}

\end{frontmatter}

\section{Introduction}
\label{sec:intro}

A current trend in flavour physics is that CP observables are approaching their Standard Model (SM) predictions.
A notable example is the $B_s^0$--$\bar{B}_s^0$ mixing phase $\phi_s$, which is expected to be very sensitive to new CP violating contributions (for a review see Ref.~\cite{Grossman:1997pa}). 
Specifically, in the SM this phase is predicted to take the tiny value $\phi_s^{\rm SM}=-2.1^\circ$, towards which the latest experimental results from the flagship $B_s \to J/\psi \phi$ analysis now appear to be converging~\cite{MPepe}:
\begin{align}
	\phi_s + \Delta\phi^\lambda_{J/\psi\phi} =
	\left\{
	\begin{array}{lcl}
		\left(0.0\pm6.0\right)^\circ & : & \textrm{LHCb~\cite{LHCb:2012conf}}, \\
		\left(-31.5^{+21.7}_{-20.6}\right)^\circ & : & \textrm{D\O\ ~\cite{Abazov:2011ry}}, \\
		\left[-34.4,6.9\right]^\circ & : & \textrm{CDF~\cite{CDF:2012ie}}.
	\end{array}
	\right.
	\label{phiSflagship}
\end{align}
Included for correctness is a possible hadronic phase shift $\Delta\phi^\lambda_{J/\psi\phi}$ for each of the final state polarizations $\lambda$~\cite{Faller:2008gt}. 
As these shifts are expected to be small, we conclude that there is no large smoking gun signal of New Physics (NP) in $B_s^0$--$\bar{B}_s^0$ mixing.

However, as precision improves, what will we be able to conclude about smallish NP in $\phi_s$?
To distinguish it from SM physics, it will become necessary to constrain the theoretical uncertainties in parameters such as $\Delta\phi^\lambda_{J/\psi\phi}$ to a sufficient accuracy.
Because these uncertainties are driven by the hadronic nature of these decays, this is not an easy task.
Standard tools include lattice QCD, QCD sum rules and SU$(3)_{\rm F}$ flavour symmetry relations between decays.
Concurrently, it would be helpful to overconstrain the phase $\phi_s$ with different analyses, analagously to the determination of the angles of the Unitarity Triangle.
In this paper we explore decay modes and analysis strategies complementary to the usual time-dependent, tagged $B_s \to J/\psi \phi$ analysis.

The layout is as follows. We first consider the optimal decay mode structure $B_s\to J/\psi s\bar s$ and how its hadronic uncertainties can be controlled.
We then discuss, in turn, two alternative final states with this structure: $B_s\to J/\psi\,\eta^{(\prime)}$ and $B_s\to J/\psi\,f_0(980)$.
Finally, we highlight the utility of effective lifetime measurements, observables that do not rely on flavour tagging, for constraining the phase $\phi_s$ and the $B_s$ decay width difference $\Delta\Gamma_s$.

\section{CP violation in $B_s\to J/\psi\,s\bar{s}$ decay modes}
\label{sec:JPsiSS}

The standard manner of measuring the CP violating phase $\phi_s$ present in the $B_s^0$--$\bar{B}_s^0$ mixing amplitude is via its interference with a specific decay amplitude for a final state $f$.
In this paper we consider only CP-eigenstates and denote the CP eigenvalue by $\eta_f$.
Because we measure the interference between mixing and decay, the CP violation of the decay mode $B_s\to f$ is also relevant and can be parameterised by the direct CP asymmetry $C_f$ and a phase shift $\Delta\phi_f$.
In practice what is measured is the time-dependent rate asymmetry, which depends on the parameters $\phi_s$, $C_f$ and $\Delta\phi_f$ as follows:
\begin{align}
	&\frac{\Gamma(B^0_s(t)\to f) - \Gamma(\bar{B}^0_s(t)\to f)}
	{\Gamma(B^0_s(t)\to f) - \Gamma(\bar{B}^0_s(t)\to f)} \notag\\
	&\quad\quad\quad = \frac{C_f\,\cos(\Delta M_s\,t) - S_f\,\sin(\Delta M_s\,t)}
	{\cosh(\Delta\Gamma_s\,t) + {\cal A}^f_{\Delta\Gamma}\sinh(\Delta\Gamma_s\,t)}
	\label{taggedRate}
\end{align}
with
\begin{equation}
	{\cal A}_{\Delta\Gamma}^f + i\,S_f = -\eta \sqrt{1-C_f^2} \exp\left[{i(\phi_s + \Delta\phi_f)}\right].
	\label{obsReln}
\end{equation}
The interfering phase is therefore the combination $\phi_s + \Delta\phi_f$ and the extraction of $\phi_s$ is limited by any theoretical uncertainty on the phase $\Delta\phi_f$.

An optimal choice for the decay mode of a $B_s$ interference measurement is a $B_s\to J/\psi s\bar s$ transition, where the $s\bar s$ quark pair forms a bound state.
This is because all the topologies contributing to this decay that contain CP violating phases are doubly Cabibbo suppressed.
Letting $\epsilon = \lambda^2/(1-\lambda^2)\simeq 5\%$ parameterise the Cabibbo suppression and explicitly extracting the CP violating phases, the $B_s\to J/\psi s\bar s$ decay amplitude in the SM has the structure:
\begin{equation}
	{\cal M}(B_s\to J/\psi\,\bar s s) \propto 1 + \epsilon\,e^{i\gamma}\,b\,e^{i\theta},
	\label{ssAmpl}
\end{equation}
where $\gamma$ is the usual angle of the CKM Unitarity Triangle.
The complex parameter $b\,e^{i\theta}$ is the ratio of CP violating over CP conserving decay topologies.
For the transition in question the CP violating topologies are all penguin in nature whereas the CP conserving topologies include, along with penguins, a colour-suppressed tree topology.
As the individual decay topologies describe hadronic, non-perturbative, processes, the parameters $b$ and $\theta$ are difficult to compute theoretically.
Fortunately, assuming the tree topology is dominant, the $\epsilon$ suppression ensures that the CP violation present in the $B_s\to J/\psi s\bar s$ is small at best: 
\begin{align}
	C_{J/\psi s\bar s}&= -2\,\epsilon\,b\,\sin\theta\sin\gamma + {\cal O}\left(\epsilon^2\right) \notag\\
	&\approx -5\%\times\left[\frac{\sin\gamma}{\sin 68^\circ}\right] \left[\frac{b}{0.5}\right]\sin\theta, \\
	\Delta\phi_{J/\psi s\bar s}&= 2\,\epsilon\,b\,\cos\theta\sin\gamma+ {\cal O}\left(\epsilon^2\right)\notag\\
	&\approx 3^\circ \times\left[\frac{\sin\gamma}{\sin 68^\circ}\right] \left[\frac{b}{0.5}\right]\cos\theta.
	\label{decayCP}
\end{align}
However, with the increasing level of precision to be offered by the LHC, these effects may no longer be negligible and it will be important to constrain the underlying hadronic physics. 

A standard tool to control the hadronic parameters $b$ and $\theta$ is via the SU$(3)_{\rm F}$ flavour symmetry of strong interactions.
In the limit that the three lightest quarks have masses that are degenerate and much lower than the $\Lambda_{\rm QCD}$ scale, the strong dynamics of the various topologies is invariant under their interchange. 
Consequently, in this limit, the hadronic parameters $b$ and $\theta$ are also common to $B_q \to J/\psi\, \bar d q$ decays for $q\in\{u,d,s\}$.
The key difference of the latter decays is that the CP violating decay topologies are not doubly Cabibbo suppressed i.e.\ the parameter $b$ is no longer accompanied by a factor of $\epsilon$:
\begin{equation}
	{\cal M}(B_q\to J/\psi\,\bar d q) \stackrel{ {\rm SU}(3)_{\rm F}}{\propto} 1 - e^{i\gamma}\,b\,e^{i\theta},
\end{equation}
These decays thereby act as {\it control channels}, allowing us to constrain the parameters $b$ and $\theta$ up to SU$(3)_{\rm F}$ breaking corrections.

A prime example is the golden decay mode $B_d\to J/\psi K_{S}$, which has been controlled using the decay $B_d\to J/\psi\pi^0$ to constrain the hadronic phase shift $\Delta\phi^d_{J/\psi K^0}$ to an accuracy of $\pm 1.5^\circ$ \cite{Faller:2008zc,Ciuchini:2011kd}.
In the future the decay mode $B_s\to J/\psi K_{S}$ should improve on this~\cite{Fleischer:1999nz,DeBruyn:2010hh}.

The question now is what to choose for the $s\bar s$ meson in the $B_s\to J/\psi\,s\bar{s}$ transition.
The vector meson $\phi$ is believed to be almost completely $s\bar s$ and is what was chosen for the flagship analysis $B_s\to J/\psi\phi$.
However, this results in a vector-vector final state that requires a complicated time-dependent angular analysis in order to resolve the CP violating phase $\phi_s + \Delta\phi^\lambda_{J/\psi\phi}$ introduced in \eqref{phiSflagship}~\cite{Dighe:1998vk,Dunietz:2000cr}.
In Ref.~\cite{Faller:2008gt} the hadronic control channels are studied for the $B_s\to J/\psi \phi$ decay.
Other bound states that are believed to contain an $s\bar s$ component are the pseudo-scalars $\eta$ and $\eta'$ and the scalar $f_0(980)$.
Their key advantage is that they do not require an angular analysis.
In the sections that follow we discuss their feasibility as alternative final states.
For a general discussion of $B_s\to J/\psi\,s\bar{s}$ modes see also Ref.~\cite{Colangelo:2010wg}.

\section{$B_s\to J/\psi\,\eta^{(\prime)}$}
\label{sec:JPsiEta}

The pseudo-scalars $\eta$ and $\eta'$ are believed to be a mixture of isospin singlet states:
\begin{align}
	|\eta\rangle &= \cos\phi_P \frac{1}{\sqrt{2}}\left(|u\bar u\rangle + |d\bar d\rangle\right) - \sin\phi_P |{s\bar s}\rangle \notag\\
	|\eta'\rangle &= \cos\phi_G \sin\phi_P \frac{1}{\sqrt{2}}\left(|u\bar u\rangle  + |d\bar d\rangle\right) \notag \\
	&\quad + \cos\phi_G\cos\phi_P |{s\bar s}\rangle + 
	\sin\phi_G |gg\rangle.
	\label{etaDefn}
\end{align}
The mixing angles are not yet tightly constrained, with $30^\circ \lesssim \phi_P \lesssim 45^\circ$ and the possible glueball state given by $|\phi_G| \sim 20^\circ$ (for details and a review see Ref.~\cite{DiDonato:2011kr}).
Because the final state is no longer purely $J/\psi s\bar s$ but now also contains $J/\psi (u\bar u, d\bar d)$ contributions dependent on the strength of the mixing angles, there are additional exchange and penguin annihilation topologies present.
Fortunately, in the SM all CP violating topologies are still doubly Cabibbo suppressed for the final state.

The $B_s\to J/\psi\,\eta^{(\prime)}$ decay modes are related by SU$(3)_{\rm F}$ flavour symmetry to the $B_d\to J/\psi\,\eta^{(\prime)}$ modes, respectively.
In Ref.~\cite{Skands:2000ru} this symmetry relation was employed in a proposed strategy to determine the CKM angle $\gamma$.
Instead, assuming we have measured $\gamma$, the $B_d\to J/\psi\,\eta^{(\prime)}$ decay modes can act as control channels for an extraction of the mixing phase $\phi_s$ from $B_s\to J/\psi\,\eta^{(\prime)}$ decays, as discussed in Section~\ref{sec:JPsiSS}~\cite{Fleischer:2011ib}.

Due to the presence of the mixing angles, however, there is in general no clean way to relate the hadronic parameters $b$ and $\theta$ of the former decays to the latter in the limit of SU$(3)_{\rm F}$ flavour symmetry.
A relation can in general be made if we may assume a hierarchy between topologies.
Namely, if we may assume that the strong amplitudes of the tree and penguin topologies contribute much more than the exchange and penguin annihilation amplitudes.
Such hierarchies can be determined by assuming flavour symmetry and considering bounds on branching ratios of decays where exchange or annihilation amplitudes are dominant.
For example, a bound on the CP violating exchange topology could be derived from an updated $B_s\to J/\psi \pi^0$ branching fraction, the current value of which dates back to an L3 measured from 1997~\cite{Acciarri:1996ur}.

In terms of experimental progress, the Belle collaboration has performed the most accurate branching ratio measurements~\cite{Belle:2012aa,Chang:2012gn}:
\begin{align*}
	{\rm BR}(B_s \to J/\psi \eta) &=  5.10^{+1.30}_{-0.97} \times 10^{-4}, \\
	{\rm BR}(B_s \to J/\psi \eta') &= 3.71^{+1.00}_{-0.85} \times 10^{-4}, \\
	{\rm BR}(B_d \to J/\psi \eta) &= 12.3^{+1.9}_{-1.8} \times 10^{-6}, \\
	{\rm BR}(B_d \to J/\psi \eta') &< 7.4 \times 10^{-6}.
\end{align*}
However, the prospects of measuring the CP observables with a time-dependent tagged analysis at the LHC are not good.
This is because $\eta$ and $\eta^{\prime}$ decay prominently to photons and neutral pions, which are difficult experimental signatures for the LHC.
Measurements beyond branching ratios will likely have to wait for the Super ``flavour'' factories.

\begin{figure}[!t]
    \centering
      \includegraphics[width=7cm]{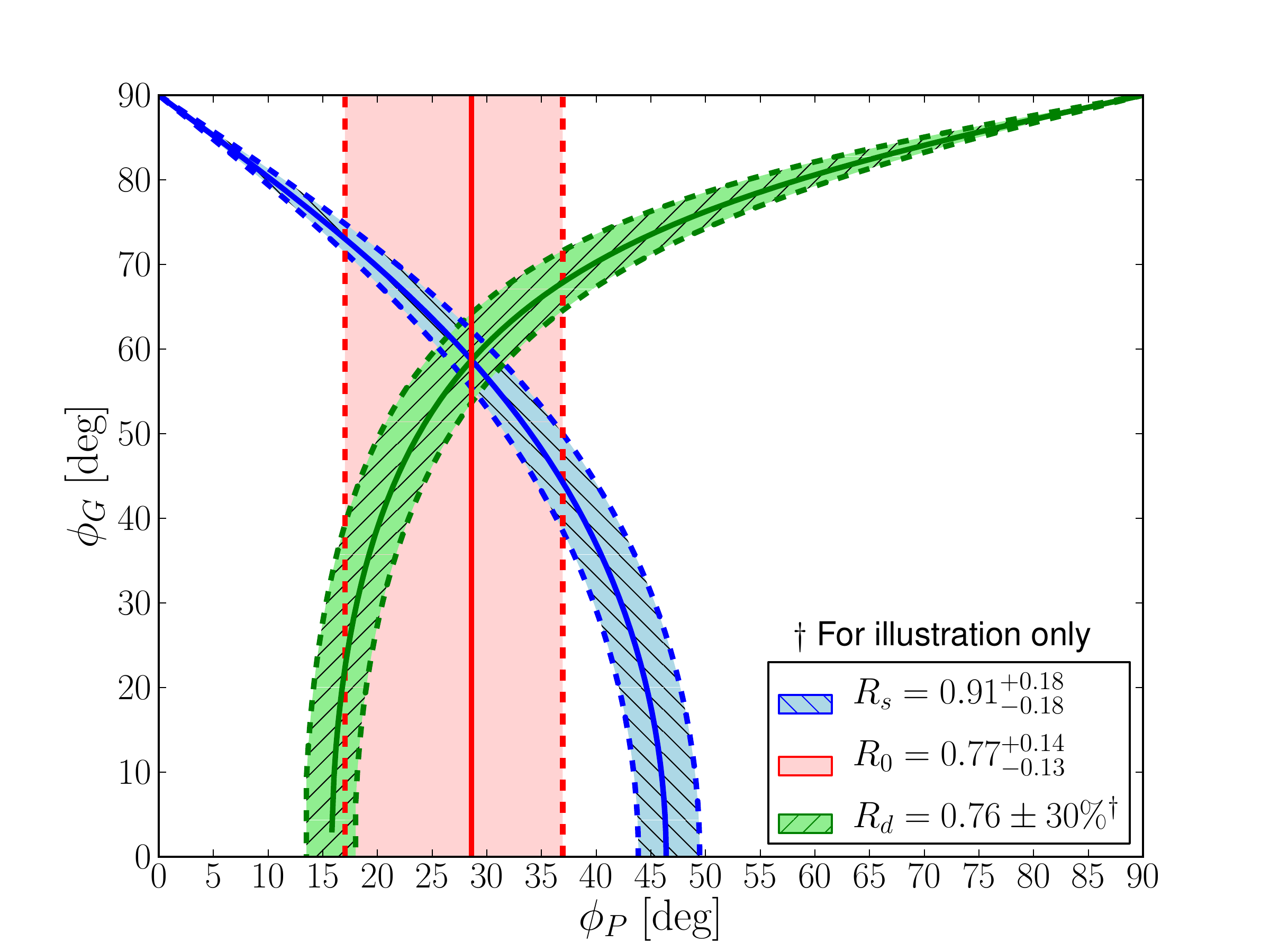} 
	  \caption{Constraints on the $\eta$-$\eta'$ mixing parameters from the $B_{s,d} \to J/\psi \eta^{(\prime)}$ and $B_d\to J/\psi\pi^0$ branching ratios, as discussed in the text. Note that not all discrete amibiguities following from the constraints in \eqref{BRconstraints} are shown.}
		 \label{fig:mixAngles}
\end{figure}

If we assume the topological hierarchy from above, it is possible to use the branching ratio measurements to constrain the $\eta$--$\eta'$ mixing angles.
To this end, we construct the ratios:~\cite{Fleischer:2011ib,Liu:2012ib}
\begin{align}
	{ R_s}&\equiv \frac{{\rm BR}(B_s \to J/\psi \eta')}{{\rm BR}(B_s \to J/\psi \eta)}\left(\frac{\Phi_s^\eta}{\Phi_s^{\eta'}}\right)^3 \notag\\
	&= \frac{\cos^2\phi_G}{\tan^2\phi_P} = 0.91\pm 0.18\notag\\
	{ R_0}&\equiv \frac{{\rm BR}(B_d \to J/\psi \eta)}{{\rm BR}(B_d \to J/\psi \pi^0)}\left(\frac{\Phi_d^\pi}{\Phi_d^{\eta}}\right)^3 \notag\\
	&= {\cos^2\phi_P} = 0.77\pm 0.14\notag\\
	{ R_d}&\equiv \frac{{\rm BR}(B_d \to J/\psi \eta')}{{\rm BR}(B_d \to J/\psi \eta)}\left(\frac{\Phi_d^\eta}{\Phi_d^{\eta'}}\right)^3 \notag\\
	&= {\cos^2\phi_G}\,{\tan^2\phi_P}
	\label{BRconstraints}
\end{align}
where $\Phi_q^P$ is the phase space factor for a $B_q\to J/\psi\,P$ transition. Similar expressions ignoring a possible glueball component ($\phi_G=0$) are given in Ref.~\cite{Datta:2001ir}.
In Figure~\ref{fig:mixAngles} we show the corresponding curves on the $\phi_G$--$\phi_P$ plane.
The contour for $R_d$ is included for illustration and has been engineered to meet at the intersection point of the other two contours.
Due to its sharp intersection with the $R_s$ contour, future measurements should allow it to strongly constrain the mixing angles.

\section{$B_s\to J/\psi\,f_0(980)$}
\label{sec:JPsi}

The decay $B_s\to J/\psi\,f_0(980)$ was originally proposed as an alternative to $B_s\to J/\psi\,\phi$ in Refs~\cite{Stone:2008ak,Stone:2009hd}.
Although its branching fraction, with $f_0(980)\to\pi^+\pi^-$, is approximately a fourth that of $B_s\to J/\psi\,\phi$, with $\phi\to K^+ K^-$, its chief advantage is that it does not require an angular analysis.
Due to the charged particles in the final state, this decay mode, unlike $B_s\to J/\psi\,\eta^{(\prime)}$, is experimentally viable at the LHC detectors. 
In fact, the LHCb experiment has reported the following measurement of its CP violating phase~\cite{LHCb:2011ab}
\begin{equation}
	\phi_s + \Delta\phi_{J/\psi f_0} = (-25\pm25)^\circ.
\end{equation}
Strictly speaking, the experiment has ignored the possible hadronic phase shift in the decay mode.
As we will now argue, the theoretical uncertainty on $\Delta\phi_{J/\psi f_0}$ is significant and should be included in the future~\cite{Fleischer:2011au}.

The underlying source of uncertainty in the CP violation present in the $B_s\to J/\psi\,f_0(980)$ transition is the uncertain nature of the $f_0(980)$.
There is no consensus in the literature on what exactly this scalar meson is and how much of it is dominated by a $s\bar{s}$ combination.
Two popular pictures of the $f_0(980)$ are that it is a conventional $q\bar{q}$ scalar meson or a tetraquark.
More exotic interpretations include a $K\bar K$ molecule (for a review see Ref.~\cite{Nakamura:2010zzi} and references therein).

In the $q\bar{q}$ description, the $f_0(980)$, together with the $\sigma(600)$, can be a mixture of isospin-singlet states with a mixing angle $\varphi_M$ in analogy to the angle $\phi_P$ for the $\eta^{(\prime)}$ states defined in \eqref{etaDefn}.
The puzzling feature of this picture, however, is why the scalars, with nonzero angular momentum, are lighter than their pseudo-scalar counterparts.

In the tetraquark picture, the $f_0(980)$ is a bound state of a diquark and an anti-diquark in a low energy configuration with no orbital angular momentum.
This interpretation reproduces the mass ordering of the light scalars in the expected way, and a mixing of less than $5^\circ$ is expected with the $\sigma(600)$ tetraquark state (which we ignore in our analysis).
In Ref.~\cite{Hooft:2008we} a consistent picture of all the scalar mesons is presented, where tetraquark and $q\bar{q}$ states mix via instanton effects.
A fit to the mass spectrum results in the light scalar mesons, with the $f_0(980)$ among them, being predominantly tetraquark.

\begin{figure}[!t]
    \centering
      \includegraphics[width=4cm]{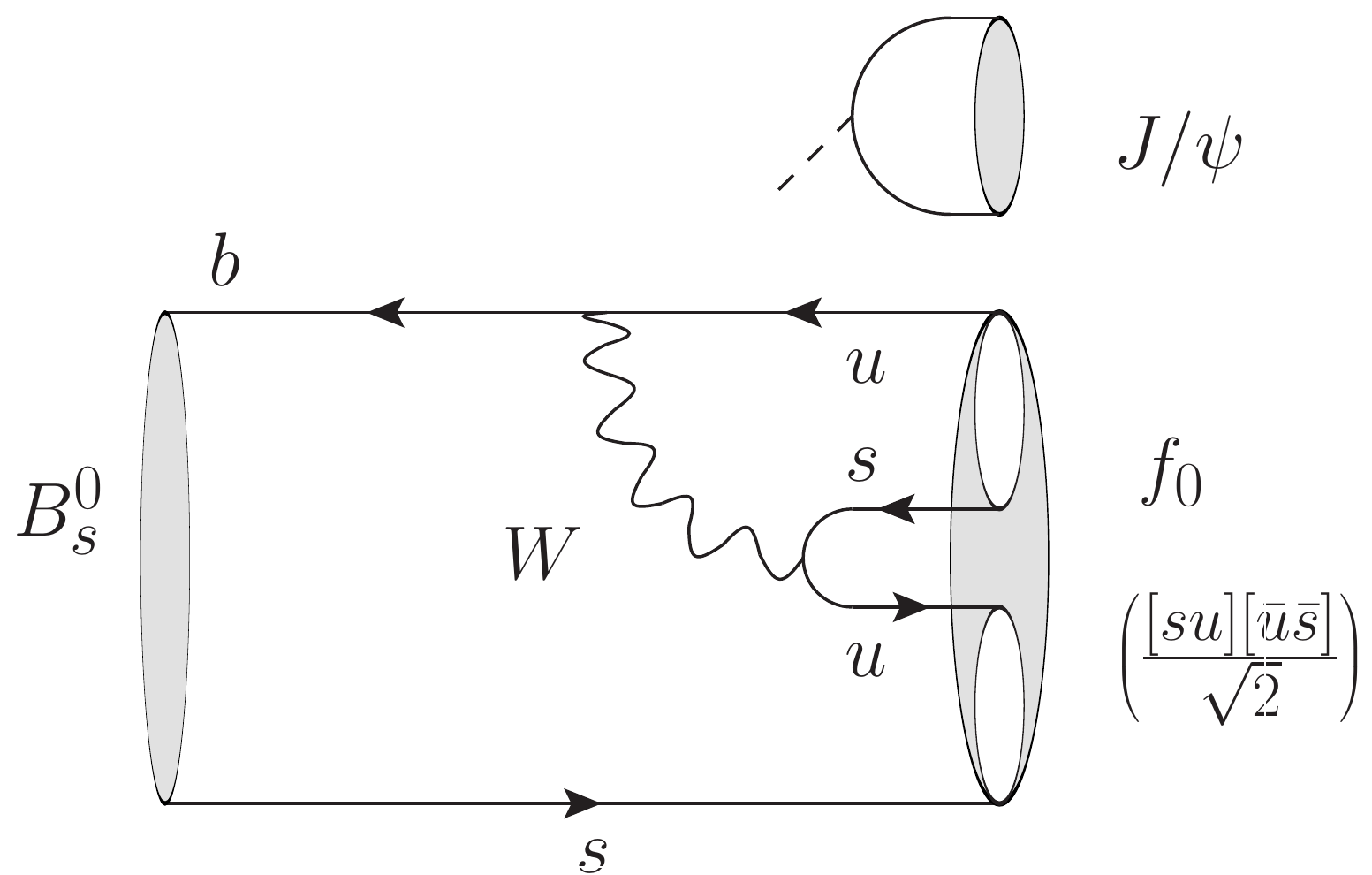} 
	  \caption{The additional topology $A_{4q}$ that contributes to the $B_s\to J/\psi\,f_0(980)$ transition in the tetraquark picture of the $f_0(980)$.}\label{fig:tetra}
\end{figure}

Which topologies contribute to the hadronic parameters $b$ and $\theta$ is strongly dependent on the $f_0(980)$ picture that one subscribes to.
In the conventional $q\bar{q}$ picture the distribution of topologies is related to the mixing angle $\varphi_M$ similarly to the $J/\psi\,\eta^{(\prime)}$ final state.
In the tetraquark picture an entirely new topology is possible, which we label $A_{4q}$ and depict in Figure~\ref{fig:tetra}, whose relative hierarchy to the other topologies is not known.

A potential control channel for constraining the hadronic parameters $b$ and $\theta$ is the transition $B_d\to J/\psi\,f_0(980)$.
However, in both pictures the dependence of the respective parameters on the various topologies in the SU$(3)_{\rm F}$ limit is sufficiently complex that we must assume a hierarchy for the topologies.
Namely, that the tree and penguin amplitudes dominate over the rest.
A possible channel to constrain the new topology $A_{4q}$ is $B_s\to J/\psi\,a_0^0(980)$ in the tetraquark picture~\cite{Fleischer:2011au}.

To estimate the current uncertainty on the CP violation in the $B_s\to J/\psi\,f_0(980)$ decay mode, we make the conservative assumption that the tree amplitude is greater than or equal to the other topologies and leave the strong phase $\theta$ unconstrained.
Setting $\gamma = (68\pm 7)^\circ$~\cite{Fleischer:2010ib}, it then follows from \eqref{decayCP} that
\begin{equation}
	{\Delta\phi_{J/\psi\,f_0}} \in [-3^\circ, 3^\circ],\quad
	\left|{C_{J/\psi\,f_0}}\right| \lesssim 0.05.
\end{equation}
Combining this with the SM prediction for the CP violating phase $\phi^{\rm SM}_s$ and inserting it into \eqref{obsReln} gives the SM prediction for the mixing-induced CP asymmetry: 
\begin{equation}
	S_{J/\psi f_0} \Big|_{\rm SM} \in \left[ -0.09,-0.01\right].	
\end{equation}
This should be compared to the na\"{i}ve SM estimate $\sin{\phi_s}|_{\rm SM}=-0.036\pm 0.002$.

The SM prediction for the branching fraction of our proposed control channel is
\begin{align}
	&{\rm BR}(B_d \to J/\psi\, f_0;f_0\to \pi^+\pi^-) \notag\\
	&\sim (1-3) \times 10^{-6} \times
	\left\{\begin{array}{ccc}
		\left[\frac{\tan\varphi_M}{\tan 35^\circ}\right]^2 &:& q\bar q\\
			1   &:&  {\rm tetraquark}
	\end{array}\right.
\end{align}
where a mixing angle of $\varphi_M= 35^\circ$ in the conventional $q\bar{q}$ picture gives an equivalent rate to the tetraquark picture.
Thus an observation of this decay would signal either a significant $d\bar d$ component for the $f_0(980)$ or hint at its tetraquark nature.

\section{$B^0_s$--$\bar B_s^0$ mixing constraints from effective lifetimes}
\label{sec:effLifetimes}

Instead of performing a full fit to the tagged time-dependent rate  for a $B_s\to f$ transition as given in \eqref{taggedRate}, it is also possible to construct observables sensitive to CP violation from the time-dependent {\it untagged} rate:
\begin{align}
	{\langle\Gamma_f\rangle} \equiv&\ \Gamma( B_s(t)\to f)\ +\ \Gamma( \bar B_s(t)\to f)\notag\\
	\propto&\ e^{-t/\tau_{B_s}}\left[ \cosh({\Delta\Gamma_s}\, t) + { {\cal A}^f_{\Delta\Gamma}} \sinh ({\Delta\Gamma_s}\,t)\right],
\end{align}
where $\Delta\Gamma_s\equiv \Gamma_{\rm L}^s - \Gamma_{\rm H}^s$ is the $B_s$ decay width difference and $\tau_{B_s}$ the mean $B_s$ lifetime.
Given enough statistics, it is possible to fit for both $\Delta\Gamma_s$ and ${\cal A}^f_{\Delta\Gamma}$ separately.
Alternatively, given limited statistics, one can instead fit a single exponential, which yields the {\it effective lifetime}. 
The effective lifetime obtained from a maximum likelihood fit to the untagged rate is analytically equivalent to the time expectation value (see the appendix of Ref.~\cite{DeBruyn:2012wj}).
We may therefore define the effective lifetime as~\cite{Fleischer:2011cw}
\begin{align}
	{ \tau_{\rm eff}} &\equiv \frac{\int_0^\infty\ t\ { \langle \Gamma_f\rangle }\ dt}{\int_0^\infty\ { \langle \Gamma_f \rangle}\ dt} \notag\\
	&=\frac{\tau_{B_s}}{1-{ y_s}^2} \left(\frac{1 +2\,{ {\cal A}^f_{\Delta\Gamma}}\, { y_s} + { y_s}^2 }
	{1+ { {\cal A}^f_{\Delta\Gamma}}\, { y_s}}\right)
	\label{effLifetime}
\end{align}
where $y_s\equiv \tau_{B_s}\Delta\Gamma_s/2$.
This observable gives us access to ${\cal A}^f_{\Delta\Gamma}$ defined in \eqref{obsReln}, which in turn depends on $\phi_s + \Delta\phi_f$ for the given final state $f$.
This dependence on ${\cal A}^f_{\Delta\Gamma}$ also makes it a key ingredient for correcting experimentally measured branching ratios in the presence of a nonzero decay with difference~\cite{DeBruyn:2012wj}. 

The most accurate determination of the  $B_s\to J/\psi\, f_0(980)$ effective lifetime is by LHCb~\cite{Aaij:2012nt}:
\begin{equation}
	\tau_{J/\psi f_0} = 1.70 \pm 0.040 \pm 0.026 \ {\rm ps}.
\end{equation}
Interestingly, this measurement deviates by more than $1\,\sigma$ from the SM expectation:
\begin{equation}
	\tau_{J/\psi f_0}\big|_{\rm SM} = (1.616\pm 0.034)\ {\rm ps},
\end{equation}
obtained from the theoretical estimate $\Delta\Gamma_s^{\rm Th} = 0.087\pm 0.021\ {\rm ps}^{-1}$~\cite{Lenz:2012mb} and $\tau_{B_s}=1.509\pm0.017$~ps~\cite{Amhis:2012bh}.

To avoid using theoretical input for $\Delta\Gamma_s$, we can instead treat it, along with $\phi_s$, as one of the parameters we seek to determine.
As we have theoretical estimates for $C_{J/\psi f_0}$, $\Delta\phi_{J/\psi f_0}$ and a measurement for $\tau_{J/\psi f_0}$, we are left with just these two unknowns.
The expression in \eqref{effLifetime}, which is a cubic equation in $\Delta\Gamma_s$, can consequently give a contour in the $\phi_s$--$\Delta\Gamma_s$ plane.
Although this by itself is already useful, the expression in \eqref{effLifetime} has another interesting feature.
Consider the decays of a $B_s$ meson to a pair of CP eigenstate final states, one CP even and one CP odd, with a comparable CP violating phase in the transition.  
Due to the dependence of ${\cal A}_{\Delta\Gamma}^f$ on $\eta_f$, as given in \eqref{obsReln}, their respective effective lifetimes will give distinctly behaving contours in the $\phi_s$--$\Delta\Gamma_s$ plane.
The intersection of these contours then pinpoints $\phi_s$ and $\Delta\Gamma_s$.

Auspiciously, the only other effective lifetime currently measured is for the $B_s\to K^+K^-$ decay mode, a CP even final state expected to have a small CP violating phase.
LHCb has the most accurate determination~\cite{Aaij:2012ns}:
\begin{equation}
	\tau_{K^+K^-} = 1.455 \pm 0.046 \pm 0.006 \ {\rm ps}.
\end{equation}
In a separate flavour symmetry analysis the CP violation of this decay was estimated using B factory data from its $U$-spin (a subgroup of SU$(3)_{\rm F}$) partner decay $B_d \to \pi^+\pi^-$, giving~\cite{1999PhLB..459..306F,Fleischer:2010ib}
\begin{equation}
	{\Delta\phi_{K^+ K^-}}= - \left(10.5^{+3.1}_{-2.8}\right)^\circ,
	\quad  {C_{K^+ K^-}}= 0.09.
\end{equation}

\begin{figure}[!t]
    \centering
      \includegraphics[width=7cm]{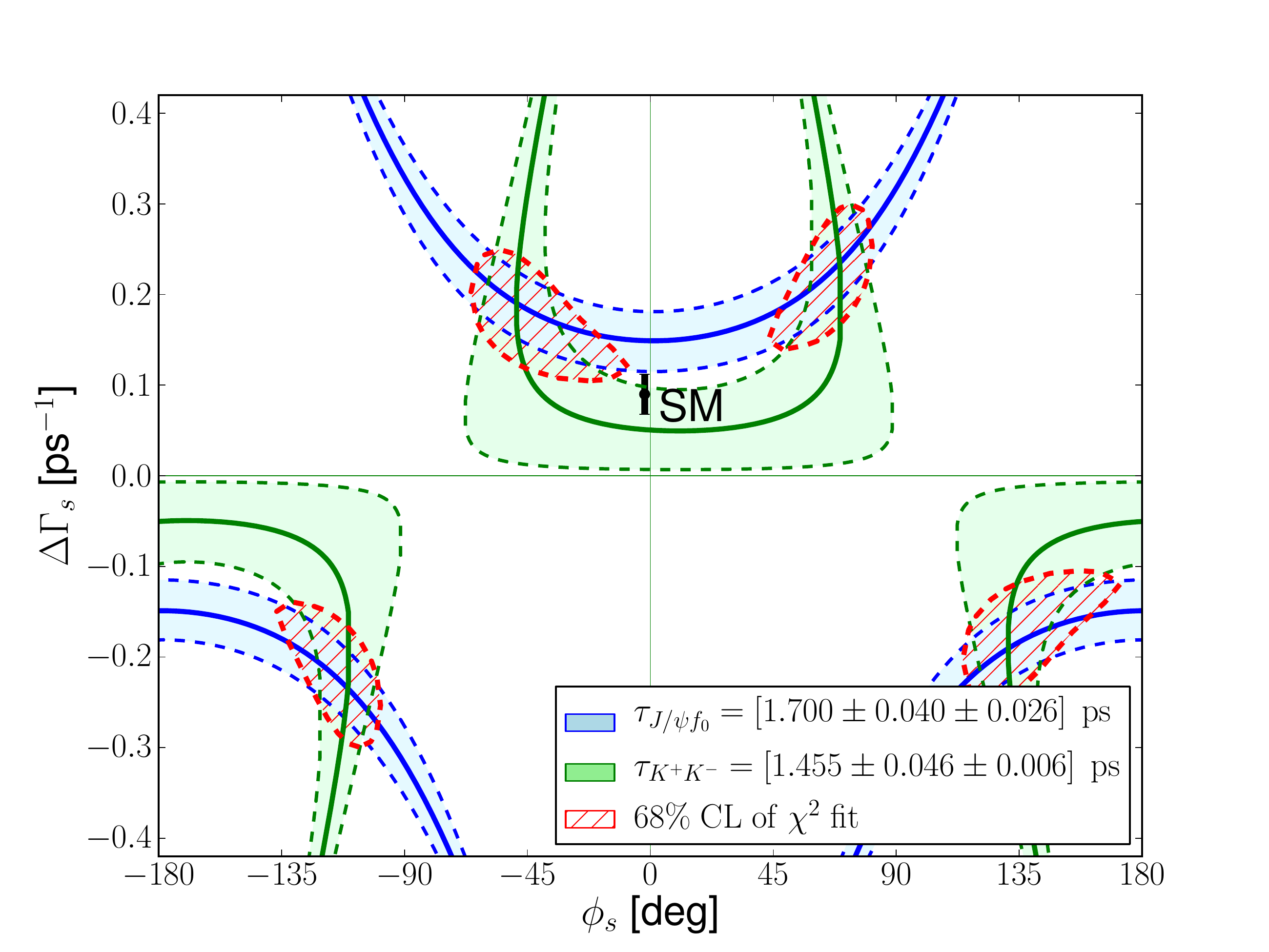} 
         \caption{The measurements of the effective $B_s\to J/\psi\,f_0(980)$ and $B_s\to K^+ K^-$ lifetimes projected onto the $\phi_s$--$\Delta\Gamma_s$ plane. The shaded bands give the 1\,$\sigma$ uncertainties of the effective lifetimes. The 68\% confidence region originates from a $\chi^2$ fit~\cite{Fleischer:2010ib}.}
		 \label{fig:DGphiS}
\end{figure}

In Figure~\ref{fig:DGphiS} we show the two contours for the CP even and odd final states on the $\phi_s$--$\Delta\Gamma_s$ plane.
The $1\,\sigma$ error bands correspond to the errors on the effective lifetime and the 68\% confidence region is the result of a combined $\chi^2$ fit.
The individual 1\,$\sigma$ confidence level results of the fit for the solution closest to the SM are~\cite{Fleischer:2011cw}
\begin{align}
	\phi_s = \left(-49^{+18}_{-11}\right)^\circ,\quad 
	\Delta\Gamma_s = 0.189^{+0.042}_{-0.052}\ {\rm ps}^{-1}, 
\end{align}
which, interestingly, do not agree well with $\phi_s=(0.0\pm6.0)^\circ$ and $\Delta\Gamma_s = (0.116\pm 0.019)\,{\rm ps}^{-1}$ coming from the flavour tagged time-dependent LHCb $B_s\to J/\psi\,\phi$ analysis~\cite{LHCb:2012conf}.
The $\phi_s$ value does agree with the less precise D\O\  and CDF results given in \eqref{phiSflagship}.

A clear advantage of the effective lifetime strategy is that it reveals the behaviour of the untagged data.
This information can be hidden in the timed-dependent tagged analysis given in \eqref{taggedRate}, where the tagged and untagged data are generally combined into a single fit.
Having multiple distinct contours pinpointing a solution in the $\phi_s$--$\Delta\Gamma_s$ plane would complement information from a global fit in a useful way, in analogy to the famous Unitarity Triangle plot.

\section*{Acknowledgements}

I would like to warmly thank the organizers of the Capri workshop on flavour physics for their invitation and superb arrangements. I would also like to thank Robert Fleischer and Giulia Ricciardi for the pleasant collaboration on the work presented here and Robert for his comments on this manuscript.

\bibliographystyle{elsarticle-num}
\bibliography{references}

\end{document}